\documentclass[twocolumn,showpacs,amsmath,amssymb,prb]{revtex4}

\usepackage{graphicx}
\usepackage{dcolumn}
\usepackage{bm}
\def\pf6{(TM\-TSF)$_2$\-PF$_6$}

\begin{document}

\title
{Comprehensive transport studies of anisotropy
and ordering phenomena in quasi-one-dimensional (TMTTF)$_2X$ salts
($X$ = PF$_6$, AsF$_6$, SbF$_6$; BF$_4$, ClO$_4$, ReO$_4$)
}
\author{B. K\"ohler}
\author{E. Rose}
\author{M. Dumm}
\author{G. Untereiner}
\author{M. Dressel}
\affiliation{1.~Physikalisches Institut, Universit{\"a}t Stuttgart,
Pfaffenwaldring 57, D-70550 Stuttgart, Germany}
\date{\today}
\begin{abstract}
The temperature dependent  dc resistivity of the
quasi-one-dimensional organic salts (TMTTF)$_2X$ ($X$ = PF$_6$,
AsF$_6$, SbF$_6$; BF$_4$, ClO$_4$, ReO$_4$) has been measured in all three crystal directions in order to investigate anisotropy, localization effects, charge and anion ordering phenomena at low temperatures. 
For all compounds and directions we extract the transport mechanisms in different regimes. The data  are thoroughly analyzed, related to structural properties, and extensively discussed in view of previous studies and latest theories.
It becomes apparent that the anions have a severe influence on the physical properties of the TMTTF salts. 
\end{abstract}

\maketitle

\section{Introduction}
For the last half a century organic conductors have been established as
models to investigate the physics in reduced
dimensions.\cite{Kagoshima88} In a one-dimensional electron gas,
Fermi-liquid theory breaks down and spin and
charge degrees of freedom become separated.\cite{Dressel03} But the
metallic phase of a solid is not stable in one dimension: as the temperature is
reduced, the electronic charges and spins tend to arrange themselves
in an orderly fashion due to strong correlations. The competition of
the different interactions upon the charge, spin and lattice degrees
of freedom is responsible for which broken-symmetry ground state is
eventually realized in a specific compound and which drives the
system towards an insulating state.\cite{Dressel07}

The family of the Fabre and Bechgaard salts have been the focus of
enormous efforts during the last three decades because small
variations of the molecules or moderate pressure tune the systems
from antiferromagnetic insulator to spin-Peierls state,
spin-density-wave state and superconductor, from a charge localized
semiconductor, charge order and ferroelectricity to a Luttinger and Fermi-liquid metal.\cite{Jerome82,Jerome94,Ishiguro98,Pouget96,Brazovskii08} These compounds are charge
transfer salts consisting of stacks of the planar organic molecules
TMTTF (which stands for tetra\-methyl\-tetra\-thia\-fulvalene) along
the $a$-axis that are separated in $c$-direction by monovalent
anions. In $b$-direction the distance of the stacks  is comparable
to the van der Waals radii. In the case of the Bechgaard salts 
the molecule TMTSF contains selenium instead of sulphur; they are mainly
metallic and become superconducting even at ambient pressure.

The understanding of the different amount of charge  localization in
the Fabre TMTTF salts and the effects of charge and anion ordering
requires reliable information on the transport properties along the
different directions. Although various groups performed numerous studies
along the stacking direction, only little is
known in the perpendicular direction. Here we present the
temperature dependent dc resistivity in all three directions of the
quasi-one-dimensional organic salts (TMTTF)$_2X$ containing anions of
octahedral ($X$ = PF$_6$, AsF$_6$, and SbF$_6$) and tetrahedral symmetry
($X$ = BF$_4$, ClO$_4$, and ReO$_4$). The findings are
compared with literature data and discussed according to latest
theory.

\section{Experimental Details}
Single crystals of (TMTTF)$_2X$ with $X$ = PF$_6$, AsF$_6$, SbF$_6$,
BF$_4$, ClO$_4$, ReO$_4$ were grown by electrochemical
methods in an H-type glass cell at room temperature. A constant
voltage of 1.5~V was applied between platinum electrodes with an
area of approximately 3~cm$^2$. The current through the solution was
between 9.2 and 13.4$~\mu$A. To reduce the diffusion, a sand barrier
was introduced. After several months we were able to harvest
needle-shaped single crystals of several millimeters in length and
less than a  millimeter in width. All compounds of the TMTTF family
are isostructural. Due the triclinic symmetry, $b^{\prime}$ denotes
the projection of the $b$ axis perpendicular to $a$, and $c^*$ is
normal to the $ab$ plane.

In order to measure the dc resistivity, small  gold contacts were
evaporated onto the natural crystal surface and thin gold wires
attached by carbon paste. Along the long $a$ axis of the crystals
and also for the $b^{\prime}$ direction four-point measurements
could be performed, while for the $c^*$ direction two contacts were
applied on opposite sides of the crystal. The samples were attached
to a sapphire plate in order to ensure good thermal contact and
slowly cooled down to helium temperatures.

\section{Results}

\begin{figure}
\centering\resizebox{0.8\columnwidth}{!}{\includegraphics*{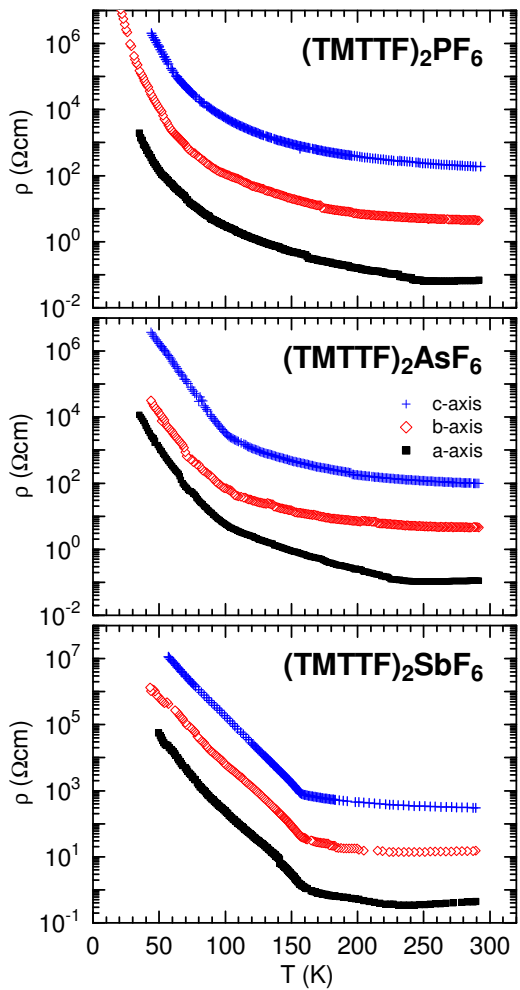}}
\caption{ \label{fig:1}(Color online) Temperature dependence of the dc resistivity in (TMTTF)$_2X$ crystals with octahedral anions $X$ = PF$_6$, AsF$_6$ and SbF$_6$ along the three crystal directions $a$, $b^{\prime}$ and $c^*$.}
\end{figure}

\begin{figure}
\centering\resizebox{0.813\columnwidth}{!}{\includegraphics*{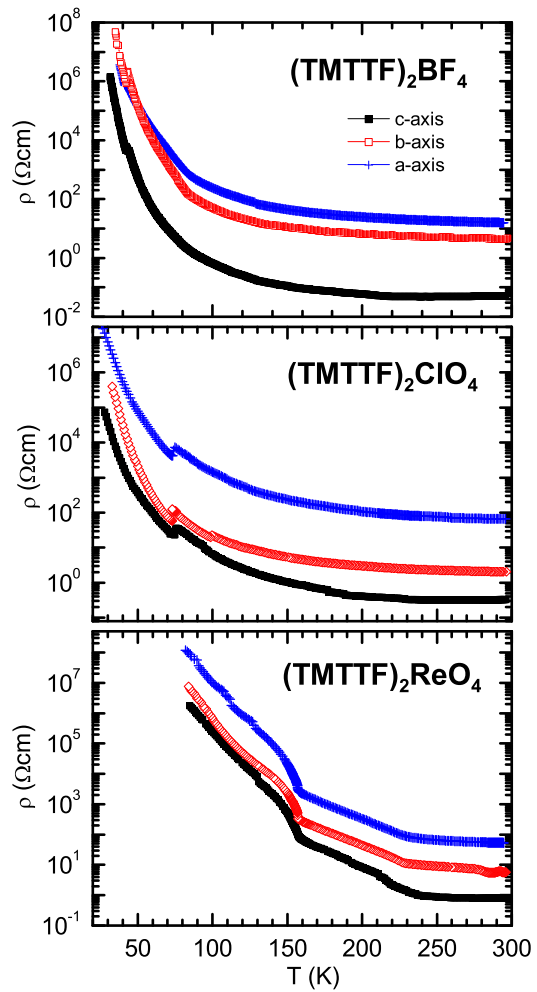}}
\caption{\label{fig:2} (Color online) Temperature dependent dc resistivity of (TMTTF)$_2X$ crystals with octahedral anions $X$ = BF$_4$, ClO$_4$ and ReO$_4$ along the three directions $a$, $b^{\prime}$ and $c^*$.}
\end{figure}

In Figure~\ref{fig:1} the temperature dependent dc resistivity of
different TMTTF salts with centrosymmetric anions  PF$_6$, AsF$_6$
and SbF$_6$ is plotted in a  logarithmic fashion to show the
overall behavior. The measurements have been performed along the
principal directions $a$, $b^{\prime}$ and $c^*$. The results for compounds with tetrahedral
anions $X$ = BF$_4$, ClO$_4$ and ReO$_4$ are presented in
Fig.~\ref{fig:2}.
The observed temperature dependences are in good accord among different batches and with previous reports.

The absolute values of the room-temperature resistivity $\rho_a$  do not differ much for the various
compounds. They range from 0.05~$\Omega$cm for (TMTTF)$_2$\-BF$_4$ to 0.8~$\Omega$cm for (TMTTF)$_2$\-ReO$_4$ and are listed in Tab.~\ref{tab:1}.
The wide spread of values reported in literature \cite{Coulon82,Moret83,Dressel01,Korin06} can be ascribed to a
number of reasons: the way contacts are attached, how the current is
injected, how the influence of micro-cracks is corrected for, and
(maybe to the least extend) the quality of the crystals. In our studies we applied a consistent procedure to all specimen that ensures a solid basis for comparison.

For very anisotropic samples it poses a challenge to measure only the contributions along the chains because strands might be broken, contain defects  or other imperfections. The one-dimensional TMTSF and TMTTF salts are generally known to be particularly susceptible to
cracks during cooling. Thus to some extend the electrical properties along the $a$ direction may be effected by $b^{\prime}$ and $c$-axes transport. Nevertheless, since the same overall behavior was repeatedly observed in numerous specimen, we could identify intrinsic properties, including similar transport mechanisms and ordering phenomena in all three directions. Applying the same procedure to all samples yields consistent results and allows for comparison between different compounds.

\subsection{Anisotropy}

It is remarkable that for all investigated (TMTTF)$_2X$ compounds the temperature dependent resistivity in many regards shows a similar temperature behavior for the three directions. The overall evolution of $\rho_a(T)$, $\rho_{b^{\prime}}(T)$ and $\rho_{c^{\ast}}(T)$ with temperature roughly differs by a constant factor ({\it i.e.}\ parallel to one another in the logarithmic representation), and the signatures of the ordering transitions are distinctly pronounced not only along the stacking direction but in $\rho_{b^{\prime}}(T)$ and $\rho_{c^{\ast}}(T)$ as well. Upon lowering the
temperature, the anisotropy in all compounds slightly decreases for $T<230~K$; {\it i.e.}\ in the temperature range where transport is in some way thermally activated.

For the three compounds with centrosymmetric anions $X$ = PF$_6$, AsF$_6$ and SbF$_6$ the anisotropy $\rho_{b^{\prime}}/\rho_a$ is approximately 30 to 50 and $\rho_{c^{*}}/\rho_a\approx 10^3$.
Notably, for the group of compounds with non-centrosymmetric anions ($X$ = BF$_4$, ClO$_4$ and ReO$_4$) the anisotropy is in general lower. In Table~\ref{tab:1} the room-temperature resistivity anisotropy
$\rho_a/\rho_{b^{\prime}}/\rho_{c^*}$ is listed for all compounds
under investigation.

\begin{table}
\caption{Room-temperature dc resistivity $\rho_a$ and anisotropy of
(TMTTF)$_2X$ for  different anions  $X$ = PF$_6$, AsF$_6$, SbF$_6$,
BF$_4$, ClO$_4$, and ReO$_4$.\label{tab:1}}
\begin{center}
\begin{tabular}{c|l|rrr}
\hline\noalign{\smallskip}
 \multicolumn{5}{c}{(TMTTF)$_2X$}\\
\noalign{\smallskip}\hline\noalign{\smallskip}
$X$&~~~$\rho_a$ &\multicolumn{3}{c}{$\rho_a$~/~$\rho_{b^{\prime}}$~~/~~$\rho_{c^*}$}\\
& ($\rm \Omega$cm) & & &\\
\noalign{\smallskip}\hline\noalign{\smallskip}
PF$_6$  &~0.08   & ~1&~~~50&2000\\
AsF$_6$ &~0.2    &1&30&1000\\
SbF$_6$ &~0.4    & 1&40&600\\
BF$_4$  &~0.05   &1&88&303\\
ClO$_4$ &~0.3    &1&4&120\\
ReO$_4$ &~0.8    &1&7&70
\\\noalign{\smallskip}\hline
\end{tabular}
\end{center}
\end{table}

\subsection{Localization and ordering phenomena}

At high temperatures basically all investigated compounds develop a metal-like behavior for transport along the $a$-direction down to a localization temperature $T_{\rho}$. In addition, the ordering phenomena -- charge order as well as anion order -- have a strong impact on the resistivity.
This results in three distinct anomalies to be identified upon cooling:
(i)~a broad minimum at elevated temperatures $T_{\rho}$ of
around 250~K which is associated  with a gradual charge localization,
(ii)~a sharp increase of $\rho(T)$ around $T_{\rm CO}=100$~K or so,
which is related to ordering of the electronic charge, and
(iii)~a kink in resistivity due to the anion ordering (AO) in the case of
tetrahedral symmetry. Details are summarized in Tab.~\ref{tab:2} and discussed in the following.
\begin{table}[h]
\caption{Transition temperatures for charge localization
$T_{\rho}$, charge order $T_{\rm CO}$, and anion order $T_{\rm
AO}$ of various Fabre salts (TMTTF)$_2X$. $T_{\rm SP}$ indicates
the spin-Peierls transition temperature.\label{tab:2}}
\begin{center}
\begin{tabular}{c|rrrr}
\hline\noalign{\smallskip}
 \multicolumn{4}{c}{(TMTTF)$_2X$}\\
\noalign{\smallskip}\hline\noalign{\smallskip}
$X$& $T_{\rho}$ & $T_{\rm CO}$ & $T_{\rm AO}$ & $T_{\rm SP}$\\
 & (K) & (K) & (K) & (K)\\
\noalign{\smallskip}\hline\noalign{\smallskip}
PF$_6$ & 250 & 67 & -~~ & 19\\
AsF$_6$& 250 & 102& -~~ & 19\\
SbF$_6$& 240 & 157& -~~ & -~ \\
BF$_4$ & 240 & 84 & 41.5& -~ \\
ClO$_4$& 260 &  -~ & 73.4& -~ \\
ReO$_4$&   -~~ & 230& 157& -~
\\\noalign{\smallskip}\hline
\end{tabular}
\end{center}
\end{table}

\paragraph*{(i) Localization}
Cooling down from room temperature, the $a$-axis resistivity
slightly decreases and passes through a broad minimum between 200~K and
300~K for all compounds except (TM\-TTF)$_2$\-ReO$_4$ which exhibits
a negative slope ${\rm d}\rho/{\rm d}T<0$ already at ambient
temperature. The minimum in resistivity is an indication of
localization effects which do not result in a sharp phase
transition but a gradual freeze out of the metallic conductivity.
However, $\rho_{b^\prime}$ and $\rho_{c^*}$ exhibits an insulating behavior in the whole investigated temperature range, implying that even at elevated temperatures there is no coherent transport present in the perpendicular directions.

\paragraph*{(ii) Charge Order}
In the temperature range between 157~K and 67~K a transition to a
charge-ordered state is observed in most (TMTTF)$_2X$ salts,
like $X$ = PF$_6$, AsF$_6$, SbF$_6$, ReO$_4$, and BF$_4$. The effect
can be barely detected in (TMTTF)$_2$PF$_6$, a distinct increase of
the slope of the resistivity is seen in the salts with $X$ =
AsF$_6$, ReO$_4$, and BF$_4$ and a obvious kink in $\rho (T)$
characterizes the data of (TMTTF)$_2$SbF$_6$ at $T_{\rm CO}$. The
transition temperatures summarized in Tab.~\ref{tab:2} are in good
agreement with published data.
\cite{Coulon82,Moret83,Pouget82,Ravy86,Nad06}
Surprisingly the ordering phenomena are not only pronounced for transport along the $a$-direction, but $\rho_{b^\prime}(T)$ and $\rho_{c*}(T)$ are strongly effected as well, indicating the three dimensional nature of the charge order (CO).

\paragraph*{(iii) Anion Order}
\begin{figure}[b]
\centering\resizebox{1\columnwidth}{!}{\includegraphics*{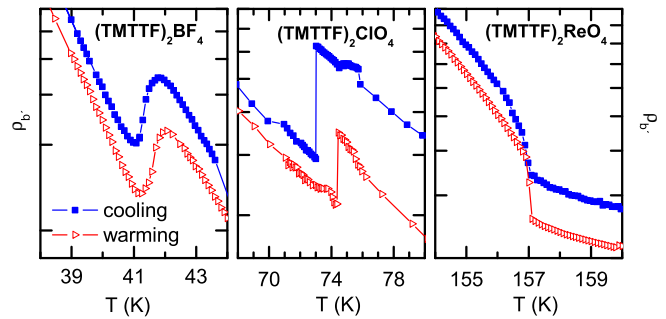}}
\caption{\label{fig:3}
(Color online) Enlarged view of the temperature dependent resistivity $\rho_{b^{\prime}}(T)$ of (TMTTF)$_2X$ for $X$ = BF$_4$, ClO$_4$ and ReO$_4$
in the range of the anion ordering. The slight shift of the transition temperature $T_{\rm AO}$ upon cooling down and
warming up evidences the first-order phase transition. For clarity reasons, the curves are vertically displaced with respect to each other.}
\end{figure}

The compounds with tetrahedral anions subsequently undergo an
anion-ordering transition at temperatures $T_{\rm AO}<T_{\rm CO}$ and the formed anion superstructure for all three investigated compounds is $q$=(1/2\,,\,1/2\,,\,1/2). In Fig.~\ref{fig:3} the resistance at the transition is depicted for the $b^{\prime}$ direction, as an example.
Upon cooling a sharp down-step in resistivity is observed for  $X$ = BF$_4$ and ClO$_4$; it is more abrupt in the latter compound.
In contrast to these two specimens, in (TMTTF)$_2$ReO$_4$ the anion order appears as a steep step-up at $T_{\rm AO}$, followed by an increase in resistivity for $T<T_{\rm AO}$.

Since this structural change of the anion order marks a first-order phase transition, a hysteretic behavior is expected. Our data show only slight  hysteresis observed when the warm-up curve is compared
to the cooling curve. In the case of BF$_4$ as anion, we find
$\delta T \approx 0.5$~K while for (TMTTF)$_2$ClO$_4$ the hysteresis
is as big as 1.5~K. For (TMTTF)$_2$ReO$_4$, no hysteretic behavior is observable, as demonstrated in Fig.~\ref{fig:3}.

The observations at the anion
order are basically identical for all three directions; this implies a strong coupling between the anions in all directions leading to three-dimensional order. It is in full accord
with structural investigations  by diffusive x-ray scattering. \cite{Moret83,Pouget82,Ravy86}

\section{Analysis}

\subsection{Transport mechanisms}

In order to analyze the temperature dependent resistivity
quantitatively and to compare it with theoretical predictions, it is
crucial to take the thermal expansion into account that is known to
be considerable for these organic charge transfer salts. From our
data $\rho^{(p)}(T)$ taken at ambient pressure $p$, the resistivity
at constant volume $\rho^{(V)}(T)$  is obtained via
\begin{equation}
\rho^{(V)}(T) = \frac{\rho^{(p)}(T)}{1+0.1 {\rm kbar}^{-1} \cdot p(T)} \quad ,
\label{eq:pressurecorrection}
\end{equation}
where $p(T)$ is the pressure required to compensate the thermal
expansion with respect to the low-temperature unit cell at $T=16$~K.
\cite{Jerome94} According to Mihaly {\em et al.} \cite{Mihaly00} and Rose {\em et al.} \cite{Rose10}
the resistivity decreases by about 10\%\ per kbar for all temperatures
above 50~K that are relevant for the present study. We assume a similar behavior for all compounds and directions.

\begin{figure}
\centering\resizebox{0.9\columnwidth}{!}{\includegraphics*{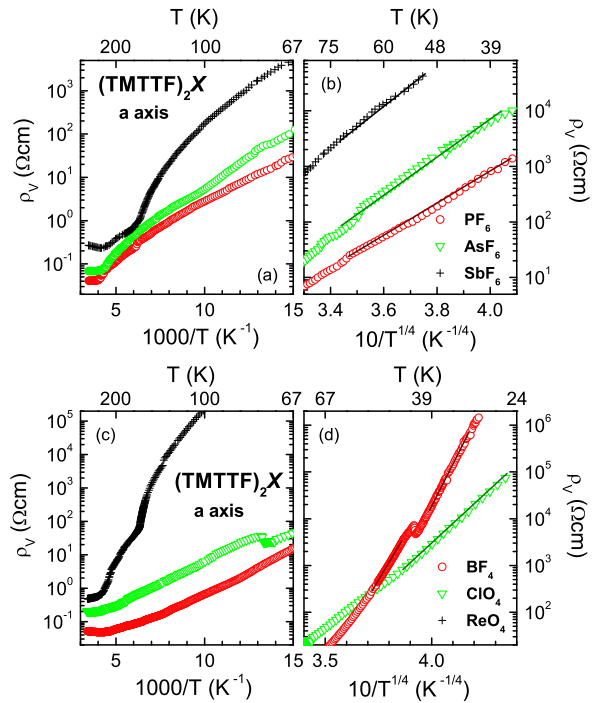}}
\caption{\label{fig:4}
(Color online) Parallel-stack dc resistivity $\rho_a^{(V)}(T)$ of various Fabre salts. The constant-pressure data are corrected according to Eq.~(\protect\ref{eq:pressurecorrection}) to account
for the thermal expansion. The upper panels (a,b) summarizes the results for the
octahedral anions: (TMTTF)$_2$PF$_6$, (TMTTF)$_2$AsF$_6$, and (TMTTF)$_2$SbF$_6$; while in the lower panels (c,d) the data for the tetrahedral anions (TMTTF)$_2$BF$_4$, (TMTTF)$_2$ClO$_4$, and (TMTTF)$_2$ReO$_4$ are shown.
On the left side (a,c) the high-temperature data are plotted as function of  $T^{-1}$ and on the right side (b,d) the low-temperature regime is presented as function of  $T^{-1/4}$. }
\end{figure}

\begin{figure}
\centering\resizebox{0.9\columnwidth}{!}{\includegraphics*{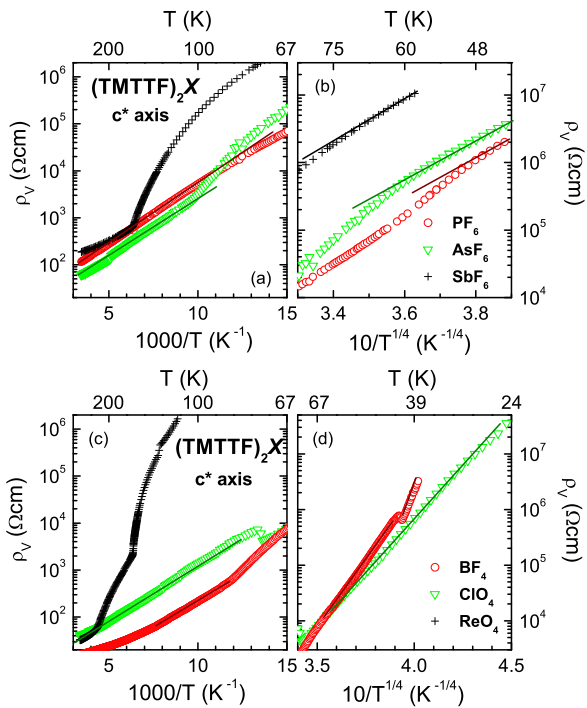}}
\caption{\label{fig:5}
(Color online) DC resistivity $\rho_c^{(V)}(T)$ along $c^*$ direction of (TMTTF)$_2$AsF$_6$, (TMTTF)$_2$PF$_6$, and (TMTTF)$_2$SbF$_6$
as function of (a) $T^{-1}$  and (b) $T^{-1/4}$. In the lower panels (c,d) the same representation for the results of (TMTTF)$_2$BF$_4$, (TMTTF)$_2$ClO$_4$, and (TMTTF)$_2$ReO$_4$.
The constant-pressure data are corrected according to Eq.~(\protect\ref{eq:pressurecorrection}) to account
for the thermal expansion. }
\end{figure}

As mentioned above, along the $b^{\prime}$ and $c^*$ directions and for $T<200$~K also along the $a$ axis, the resistivity always increases with
decreasing temperature. Different temperature-dependent transport
mechanisms can account for such a semiconducting behavior. For band
transport, the resistivity follows the Arrhenius law
\begin{equation}
\rho_{\rm dc}^{(V)}(T) = \rho_0 \exp\left\{\frac{\Delta}{T}\right\}\quad . \label{eq:arrhenius}
\end{equation}
If the energy gap $\Delta$ is temperature independent, the logarithm
of the resistivity should be linear in $1/T$. The left panels of
Figures~\ref{fig:4} and ~\ref{fig:5} show that this is only the case
for transport along the $c^*$-axis at
elevated temperatures above the ordering transitions. Along the stacking direction, we cannot identify a
sufficiently large range of simple thermally activated transport according to Eq.~(\ref{eq:arrhenius}).

An alternative transport process is phonon-activated hopping
between localized states which in disordered materials is commonly
described by the variable-range-hop\-ping model proposed by Mott:
\cite{Mott79}
\begin{equation}
\rho_{dc} \propto \exp\left\{\left(\frac{T_0}{T}\right)^{1/\gamma} \right\} \quad ,
\label{eq:Mott}
\end{equation}
where $\gamma=d+1$ is related to the dimension $d$ of the system. The
right panels of Figs.~\ref{fig:4} and ~\ref{fig:5} indicate that the
transport in all compounds is governed by three-dimensional hopping
below approximately $T=60$~K. As expected from the very similar over-all behavior, the quasi-one-dimensional
nature of the material does not influence this incoherent transport process: $d=3$; the anisotropy only enters the prefactor, but not the power law.

\subsection{Temperature evolution of the energy gap}
\begin{figure}
\centering\resizebox{\columnwidth}{!}{\includegraphics*{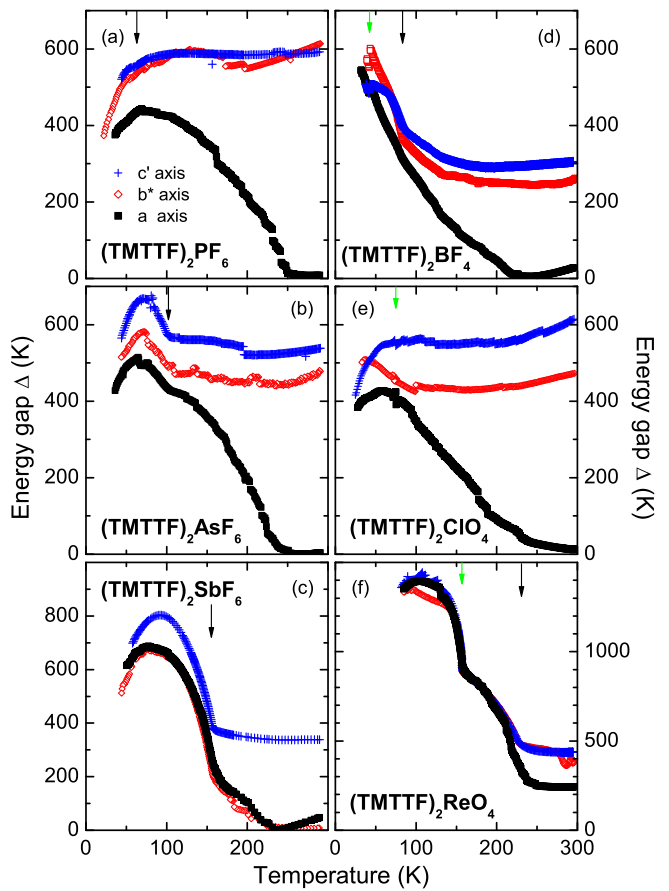}}
\caption{\label{fig:6}
(Color online) Energy gaps of (TMTTF)$_2X$ as function of temperature along $a$, $b^{\prime}$ and $c^*$ axes. The activation energy was determined
from the constant volume data of the resistivity according to Eq.~(\ref{eq:gapT}) choosing an appropriate value of $\rho(T)$. The small green arrows indicate $T_{\rm AO}$, the larger black arrows denote the charge ordering temperature $T_{\rm CO}$.}
\end{figure}
Along the $a$ axis neither of the models can satisfactorily fit the data at higher temperatures. If the current is
applied parallel to the $c^{*}$ axis, on the other hand,
both models fail in the regions
below the charge-order transitions and also the anion-order
transition in (TMTTF)$_2$ReO$_4$. Hence, we abandon the assumption of
a constant activation energy $\Delta$ in Eq.~(\ref{eq:arrhenius}).
Since the charge-order transition
is a second-order phase transition, we expect
a mean-field temperature dependence of the energy gap. In
Fig.~\ref{fig:6} the temperature-dependent energy gap
\begin{equation}
\Delta(T) = \ln\left(\frac{\rho^{(V)}(T)}{\rho_0}\right)\cdot T    \label{eq:gapT} 
\end{equation}
is plotted as function of $T$ for all compounds
and directions investigated.\cite{remark1} It immediately becomes obvious that the observed behavior is more complex and require a
detailed discussion.

\begin{figure}
\centering\resizebox{0.7\columnwidth}{!}{\includegraphics*{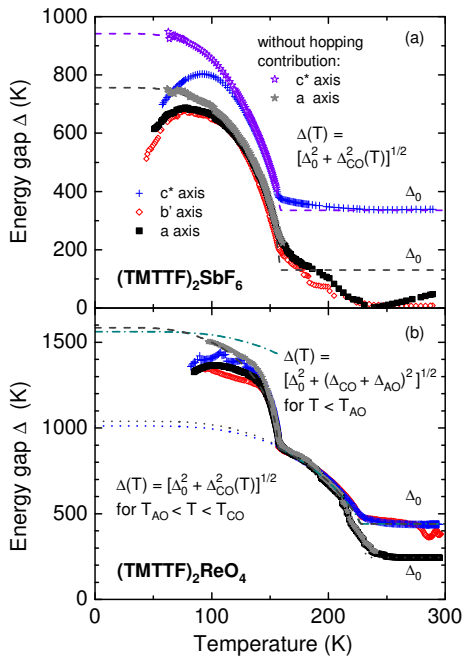}}
\caption{\label{fig:7}
(Color online) (a) Energy gaps of (TMTTF)$_2$SbF$_6$ as function of temperature along $a$, $b^{\prime}$, $c^*$ axes. The contribution by hopping conduction to the transport [ $\rho_{c^*} ^{\rm hop}(T)= (4.31\times10^{-5}~\Omega {\rm cm})\exp\{72.45~{\rm K}/T^{1/4}\}$, $\rho_{a} ^{\rm hop}(T)= (39.8\times10^{-3}~\Omega {\rm cm})\exp\{74.5~{\rm K}/T^{1/4}\}$]  was subtracted and the remaining curve fitted by Eq.~(\ref{eq:gapquadrat}) (dashed line), shown exemplary here only for two crystal directions.
(b)~To eliminate the hopping conduction for (TMTTF)$_2$ReO$_4$, the dotted  curves are obtained by a fit with Eq.~(\ref{eq:gapquadrat}). Below the AO transition, an additional gap opens and the complete behavior follows Eq.~(\ref{eq:gapquadrat+AO}) (grey dashed line).  The fit for a temperature independent gap $\Delta_{\rm AO}$ is indicated by the dashed-dotted curve 
(cyan).
Activation energies obtained from the fits are listed in Tab.~\ref{tab:3} for the limit $T\rightarrow 0$.}
\end{figure}

Basically for all curves plotted in Fig.~\ref{fig:6}, the energy
gap reaches a maximum well below $T_{\rm CO}$, and starts to
decrease for even lower temperatures. In general, this is an
indication of two processes with different temperature
dependences. From the fit to Eq.~(\ref{eq:Mott}) presented in
panels (b) and (d) of Figures~\ref{fig:4} and \ref{fig:5}, we know
that in this low-temperature range the hopping of charge carriers
between localized states starts to contribute to the electronic
transport. This second contribution taints the analysis of
$\rho(T)$ in terms of an Arrhenius law and cannot be adjusted for
by a temperature dependent gap. We therefore determine the hopping conductivity
in the low-temperature limit and subtract this contribution from
the total conductivity in order to arrive at a  purely activated
behavior. As seen  in Fig.~\ref{fig:7}(a) for the example of
(TMTTF)$_2$SbF$_6$ and (TMTTF)$_2$ReO$_4$, now the energy gap
monotonically increases with decreasing temperature and finally
levels off.

Obviously two regions can now be distinguished in the curves of Fig.~\ref{fig:7}.
At high temperatures the energy gap is more or less temperature independent with a value $\Delta_0$ of a few hundred Kelvin. The reason for this high-temperature energy gap is still under discussion, but commonly related to the bond dimerization, resulting in an unequal distribution of the charge between the molecules.
Cooling through $T_{\rm CO}$, the electronic charge on the molecular
sites becomes disproportionate due to charge order. The charge-order gap
exhibits a BCS-like increase with lowering the temperature and
amounts up to approximately 1000~K. Qualitatively the same behavior
(at $T < T_{\rm CO}$) is observed for all specimen and directions.
The total energy gap consisting of these two contributions is well described by
\begin{equation}
\Delta(T) = \sqrt{\Delta_0^2+\Delta_{\rm CO}^2(T)}
\quad , \label{eq:gapquadrat}
\end{equation}
following the idea that these two effects cause a spatial modulation of the potential, however, displaced by a quarter of a wavelength.
Only in some cases the energy gap seems to be slightly
temperature dependent above $T_{\rm CO}$, which prevents an unambiguous decomposition
and satisfactory description.

In the case of tetrahedral anions, a second transition occurs at $T_{\rm AO}$, causing a further energy gap $\Delta_{\rm AO}$ to open in most of the (TMTTF)$_2$X compounds. As demonstrated in Fig.~\ref{fig:7}(b) for the example of (TMTTF)$_2$ReO$_4$, it can be  described best by an additional temperature dependent (BCS-like) energy gap. This result surprises considering the nature of the anion order transition. Usually structural transitions are of first order, and in (TMTTF)$_2$ReO$_4$ the anion order is attended by the doubling of the unit cell along the $a$-direction. Hence, intuitively the expectations are a temperature independent gap $\Delta_{\rm AO}$, but this does not allow for a satisfactory description [dash-dotted line in Fig.~\ref{fig:7}(b)] of the results. 
The temperature dependence of $\Delta_{\rm AO}$ infers that 
the gap is not just caused directly by the anion potential, 
but the main contribution comes via changes in the 
TMTTF stacks triggered by the anion ordering; 
e.g.\ modifications of the on- and off-side potentials. 
These changes result in different charge concentrations on and between the organic molecules and alter the back-scattering potentials felt by the charged particles taking part in transport.  
If one takes this mechanism in account, the spatial modulation of the charge-order effect and the resulting changes caused by the anion order are in phase. The total energy gap is then described by:
\begin{equation}
\Delta(T) = \sqrt{\Delta_0^2+[\Delta_{\rm CO}(T)+\Delta_{\rm
AO}(T)]^2} \quad . \label{eq:gapquadrat+AO}
\end{equation}

Let us now consider the various compounds in more detail to point out their peculiarities.

\paragraph*{(i) {\rm (TMTTF)$_2$$X$ ($X$=PF$_6$, AsF$_6$, SbF$_6$)}}
The charge-ordering transition in (TMTTF)$_2$PF$_6$ at $T_{\rm
CO}=67$~K leads only to a small increase of the energy gap along
all three directions. It can be barely seen in Fig.~\ref{fig:6}(a)
because at that temperature the hopping dominates already the
transport behavior and the transition seems to be smeared out.
For (TMTTF)$_2$AsF$_6$ a mean-field-like increase of the energy gap
of about 110 K is seen at $T_{\rm CO}=103$~K along  all
three directions. Below 75~K, hopping transport dominates.
In (TMTTF)$_2$SbF$_6$ the effects at the CO transition 
($T_{\rm CO}=157$~K) are
most pronounced. Along all three axes the energy gap increases by
several hundred Kelvins. The temperature dependence has a BCS 
shape. As demonstrated in Fig.~\ref{fig:7} the high-temperature
gap and the charge-order gap add in quadrature
[Eq.~(\ref{eq:gapquadrat})]. If the contribution of the hopping
transport to the conductivity is subtracted, the mean-field-like
temperature dependence of the gap can be extracted to lowest
temperatures with $\Delta_{\rm CO}\approx 800$~K. No indications of
a first-order nature of the transition are found.

\paragraph*{(ii) {\rm (TMTTF)$_2$ClO$_4$}}
No $2k_F$ charge order has been observed in this compound.
At the anion-order transition ($T_{\rm AO}=73$~K) of
(TM\-TTF)$_2$\-ClO$_4$, the resistivity shows a sudden decrease
(Fig.~\ref{fig:3}). However, from Fig.~\ref{fig:6}(e) we see that
the slope in the Arrhenius plot is unaffected by the jump, giving
evidence that $\Delta$ is constant across the transition. We
therefore attribute the abrupt change in the resistivity to the
prefactor $\rho_0$ in Eq.~(\ref{eq:arrhenius}). From
Fig.~\ref{fig:2} it becomes obvious that at $T_{\rm AO}$ 
$\rho_0$ decreases by the
same factor of 2.5 for all three directions.

\paragraph*{(iii) {\rm (TMTTF)$_2$BF$_4$}}
In the case of (TMTTF)$_2$BF$_4$, the jump in $\rho(T)$ at $T_{\rm
AO} = 42$~K is not so abrupt as in the ClO$_4$ counterpart. The
effect is strongest along the $b^{\prime}$ and weakest along the
$c^*$ axis. A close inspection of the Arrhenius plot reveals a weak
increase  in the slope ${\rm d}\rho/{\rm d}T$ for $T<T_{\rm AO}$.
At these low temperatures, however, substantial contributions from
hopping transport might be present and  obscure a precise
analysis. Nevertheless, we estimate that $\rho_0$ decreases by
about the same ratio as in (TMTTF)$_2$ClO$_4$, while the energy
gap $\Delta(T)$ increases slightly as the temperature passes
through the anion ordering transition upon cooling.
At $T_{\rm CO}=84$~K the energy gap increases in a mean-field
manner due to the opening of a charge-order gap. The rise is
moderate and best seen along the $c^*$ axis. The effect is diluted
by the hopping transport which gets dominant in this temperature
range.

\paragraph*{(iv) {\rm (TMTTF)$_2$ReO$_4$}}
In this compound, strong contributions to the gap are observed by
the charge and by the anion ordering. When cooling through the CO
transition ($T_{\rm CO}=223$~K), the energy gap $\Delta(T)$
increases by several hundred Kelvins, in a similar fashion for all
three axes. As demonstrated in Fig.~\ref{fig:7}(b), the
temperature dependence of the gap resembles  a BCS-like shape.
Following Eq.~(\ref{eq:gapquadrat}), the total energy gap is
a result of contribution due to temperature independent charge localization and charge order adding up in quadrature.
As mentioned above, (TMTTF)$_2$ReO$_4$ exhibits a completely
different behavior of the conductivity at $T_{\rm AO}$, where the increase in $\rho(T)$ can be described by Eq.~(\ref{eq:gapquadrat+AO})
The enhancement of the gap is
not abrupt, but stronger than expected for a second-order
transition. In the low-temperature limit the extra contribution to the activation energy in the anion-ordered state is very similar for all three directions. One reason of the particular behavior of (TMTTF)$_2$ReO$_4$ -- and similar (TMTTF)$_2$FSO$_3$ -- is the large size of the anions which modifies the array of the TMTTF molecules. \cite{Takahashi06}
\\

In Table~\ref{tab:3} we list the activation energies $\Delta$
obtained  from our analysis of $\rho(T)$ for all the compounds
investigated; in addition we give the individual contributions of
the charge localization $\Delta_0$, the charge order $\Delta_{\rm
CO}$ and the anion order  $\Delta_{\rm AO}$.
\begin{table}
\caption{Summary of the activation energies obtained from
Fig.~\ref{fig:6} in the three directions of the various (TMTTF)$_2X$ salts.
In the localization regime $T_{\rho}>T>T_{\rm CO}$ an energy gap $\Delta_0$ is dominant. The charge-ordered regime $T_{\rm
CO}>T>T_{\rm AO}$ is governed by $\Delta_{\rm CO}$; below the
anion-ordering-transition temperature $T_{\rm AO}>T$ the energy gap $\Delta_{\rm
AO}$ gives an additional contribution in the case of (TMTTF)$_2$ReO$_4$.
$\Delta$ is the total energy gap in the limit $T \rightarrow 0$~K. $\vartheta(T)$ considers some additional temperature dependence; uncertain cases are indicated by parentheses; sometimes only very rough estimates are possible. \label{tab:3}}
\begin{center}
\begin{tabular}{c|cccc}
\hline\noalign{\smallskip}
 \multicolumn{5}{c}{(TMTTF)$_2X$}\\
\noalign{\smallskip}\hline\noalign{\smallskip}
direction   & $\Delta_0$ (K) & $\Delta_{\rm CO}$ (K)& $\Delta_{\rm AO}$ (K) & $\Delta$ (K)\\
\noalign{\smallskip}\hline\noalign{\smallskip}
&\multicolumn{4}{c}{$X$ = PF$_6$}\\
\noalign{\smallskip}\hline\noalign{\smallskip}
$a$         & $440\pm \vartheta(T)$ & small  & - & $440\pm 20$\\
$b^{\prime}$ & $590\pm 30$ & small  & - & $590\pm 30$\\
$c^*$       & $590\pm 10$ & small  & - &  $590\pm 10$\\
\noalign{\smallskip}\hline\noalign{\smallskip}
&\multicolumn{4}{c}{$X$ = AsF$_6$}\\
\noalign{\smallskip}\hline\noalign{\smallskip}
$a$         & $420\pm \vartheta(T)$ &  310 & - & 520 \\
$b^{\prime}$& $460\pm 20$ &  370  & - & 590 \\
$c^*$       & $540\pm 20$ &  410  & - & 680 \\
\noalign{\smallskip}\hline\noalign{\smallskip}
&\multicolumn{4}{c}{$X$ = SbF$_6$}\\
\noalign{\smallskip}\hline\noalign{\smallskip}
$a$         & $130\pm \vartheta(T)$ & $ 745\pm 20$ & - & $760 \pm 30$\\
$b^{\prime}$& $130\pm \vartheta(T)$ & $ 745\pm 20$  & - & $760 \pm 30$\\
$c^*$       & $340\pm 20$  & $ 880\pm 20$  & - & $940 \pm 30$ \\
\noalign{\smallskip}\hline\noalign{\smallskip}
&\multicolumn{4}{c}{$X$ = BF$_4$}\\
\noalign{\smallskip}\hline\noalign{\smallskip}
$a$         & $390\pm \vartheta(T)$  &  (medium) & small  & ($610$) \\
$b^{\prime}$& ($400\pm \vartheta(T)$) &  medium  & small & ($600
$)\\
$c^*$       & $500\pm \vartheta(T)$ &  310  & small & $ 590$ \\
\noalign{\smallskip}\hline\noalign{\smallskip}
&\multicolumn{4}{c}{$X$ = ClO$_4$}\\
\noalign{\smallskip}\hline\noalign{\smallskip}
$a$         & $430\pm \vartheta(T)$ & - & 0 & $ 430\pm 50$ \\
$b^{\prime}$& ($440\pm 60$) & - & 0 & $ (440\pm 60$) \\
$c^*$       & $560\pm 50$  & - & $0$ & $560\pm 50$ \\
\noalign{\smallskip}\hline\noalign{\smallskip}
&\multicolumn{4}{c}{$X$ = ReO$_4$}\\
\noalign{\smallskip}\hline\noalign{\smallskip}
$a$         & $245\pm 50$ & $ 1010\pm 50$ & $ 530\pm 20$ & $1560 \pm 60$\\
$b^{\prime}$& $440\pm 70$ & $ 910\pm 50$  & $ 590\pm 20$ & $1560 \pm 80$\\
$c^*$       & $440\pm 50$ & $ 910\pm 50$  & $ 590 \pm 20$ & $1560 \pm 60$\\
\noalign{\smallskip}\hline
\end{tabular}
\end{center}  
\end{table}
Only (TMTTF)$_2$AsF$_6$ has previously been analyzed along all three
directions;\cite{Korin06} the published activation energies agree
very well with the present findings. Nad and Monceau \cite{Nad06}
measured the dielectric response in stacking direction and were able
to extract activation energies for several of TMTTF compounds;
except for the AsF$_6$ and PF$_6$ salts, their values are good
accord with our results.

\subsection{Anisotropic transport}

If the temperature is low enough, no coherent transport exists in (TMTTF)$_2X$ compounds along any of the three axes; instead hopping transport is observed in the $a$, $b^{\prime}$ and $c^*$ directions.\cite{remark4}
At high temperatures, the situation is not that clear and well resolved.
Along the $b^{\prime}$ axis, a more or less temperature independent energy gap is found in all compounds but (TMTTF)$_2$SbF$_6$ and (TMTTF)$_2$ReO$_4$. The values are in the same range
or up to 100~K smaller than in $c^*$ direction.

\paragraph*{(i) $a$-axis}
The $a$-axis resistivity goes through a minimum in most compounds
which corresponds to a zero energy gap, as seen in
Fig.~\ref{fig:6}. Then with decreasing $T$ the gap increases up
to 400 K. There are some abrupt changes in the temperature
dependent gap due to the ordering transitions.

\paragraph*{(ii) $c^{\ast}$-axis}
For $T>100$~K and above any ordering temperature, the charge transport along the $c^*$ direction is  thermally activated with a corresponding
energy of approximately $540 \pm 40$~K. Following
Georges {\it et al.}\cite{Georges00}
we conclude that the $c^{*}$-axis dc data evidence
thermally activated behavior with an activation energy equal to
the charge gap $\Delta_0$ that is very similar for all TMTTF salts.
Somewhat smaller values are observed in the
compounds with higher ordering temperatures, like (TMTTF)$_2$ReO$_4$
($\Delta_0=440$~K) and (TMTTF)$_2$SbF$_6$ ($\Delta_0=340$~K). In these
compounds the results might be influenced by the high transition temperatures and fluctuations.

At low temperatures ($T< 70$~K and not to close to the CO
transition) the resistivity exhibits the Mott  $T^{1/4}$ law for
hopping transport [Eq.~(\ref{eq:Mott})]. As demonstrated in
Fig.~\ref{fig:7}, the low-$T$ conductivity can be described very
well by the sum of an Arrhenius and a $T^{1/4}$ behavior. These
are two completely separate concepts of electronic transport which
seem to exclude each other. On the one hand our experimental
findings of an Arrhenius law suggest band transport above 100 K
while hopping transport is dominant well below that temperature.
On the other hand, it is known that in the TMTTF salts the
molecular chains are well separated along the $c$ axis and that the
hopping integral $t_c$ is rather small (less than 1~meV) which
does not allow for any considerable overlap of the molecular orbitals.

\paragraph*{(iii) $b^{\prime}$-axis}
Along the $b^{\prime}$ axis the same arguments  apply
as along the $c^{*}$ direction, however, the hopping integrals (orbital
overlap) and thus the absolute values of conductivity are
considerably larger. We expect the energy  gaps derived from the
Arrhenius law being temperature independent and isotropic for
the origin is always the effective Coulomb repulsion on the
chains.  Hence, the resulting  resistivity exhibits the same
temperature dependence in all directions, assuming that the
hopping integrals are temperature independent and no other
contributions to the charge transport are present. Our results
suggest that this condition is fulfilled best along the $c^{*}$ axis.

\section{Discussion}

According to band-structure calculations, the (TMTTF)$_2X$ compounds develop bands along the $a$ direction which are $1/4$ filled by chemistry. Due to a slight dimerization of the chains, the conduction band is split and effectively half filled. Thus one would expect coherent metallic transport along the chains, and this agrees with $\rho_a(T)$ we observe at elevated temperatures.
On-site Coulomb repulsion $U$ splits the conduction band in upper and lower Hubbard bands by a Mott gap $\Delta_0$. Correspondingly, below $T_{\rho}\approx 250$~K the transport becomes thermally activated according to  Eq.~(\ref{eq:arrhenius}). As charge order develops for $T<T_{\rm CO}$ the gap enlarges upon cooling. In the case of anion ordering, its size increases even more in most compounds.

The small transfer integrals ($t_b\approx 10$~meV and $t_c\approx 1$~meV) prevent coherent charge transport in the perpendicular directions, and one would expect diffusive motion only.
In fact, insulating behavior is observed in the perpendicular directions at all temperatures.
However, variable range hopping causes a very distinct temperature dependence, given in Eq.~(\ref{eq:Mott}), for instance, which could be identified only at very low temperatures (right panels of Fig.~\ref{fig:5}).
The intriguing question is why most of the time the perpendicular transport basically follows the behavior parallel to the chains at reduced temperatures, except for some factor. This immediately implies that diffusive transport does not influence the temperature dependence of $\rho_{b^{\prime}}$ and $\rho_{c^*}$, and that $t_{\perp}$ mainly contributes to the prefactor. Thermally activated-transport is the dominant mechanism also perpendicular to the stacks. Even when tunneling between the chains, the charge carriers have to overcome the energy gaps that open in the conduction band due to Coulomb repulsion, charge order and anion order. The behavior of $\rho_{b^{\prime}}(T)$ and $\rho_{c^*}(T)$ reflects the changes of the one dimensional band.

Within the group of symmetric and non-symmetric anions there are trivial relations of the absolute values $\rho(300~{\rm K})$ in the three directions to the corresponding lattice parameter,summarized in footnote~\onlinecite{remark5} and Fig.~\ref{fig:11}. 
The absolute value of $\rho(300~{\rm K})$ depends on the orbital overlap defined by distance and position of the organic molecules with respect to each other. Along the $a$-axis the transfer integrals determine the bandwidth, along the two other crystal directions they affect the tunnelling probability. This description does not include any correlation effects that might have impact even on the room temperature resistivity, e.g. 
fluctuations founded by Coulomb repulsion or rotating anions causing localization effects.
Transfer integrals and correlation effects can be changed by chemical or physical pressure.
A detailed investigation of the pressure dependence of these charge localization and ordering effects will be given elsewhere (Ref.~\onlinecite{Rose10}).

\subsection{Anisotropic transport mechanism}

\begin{figure}
\centering\resizebox{0.7\columnwidth}{!}{\includegraphics*{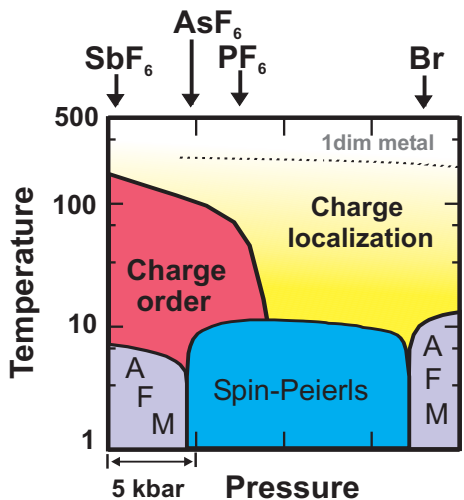}}
\caption{\label{fig:8} (Color online) The phase diagram of (TMTTF)$_2X$ shows the different phases (antiferromagnetic AFM, spin-Peierls, charge ordered, charge localization, one-dimensional metal) as a function of pressure. While some of the boundaries are clear
phase transitions, the ones indicated by dashed lines are better
characterized as a crossover. The position in the phase diagram
can be tuned by chemical pressure or external pressure which the approximate scale given. For the different compounds with octahedral and spherical anions,
the ambient-pressure position in the phase diagram is indicated by the arrows on the top axis.}
\end{figure}

There have been multiple attempts to describe the transport in anisotropic conductors theoretically. Ishiguro {\it et al.}\cite{Ishiguro80} considered one-dimensional metallic chains connected by diffusive conduction and suggested that the ratio of the conductivity parallel and perpendicular to the stacks is
\begin{equation}
\label{eq:7}
\frac{\sigma_{\parallel}}{\sigma_{\perp}}=\frac{\hbar^2 v_F^2}{2 b^2 t_{\perp}^2} \quad ,
\end{equation}
where $v_{F}$ is the Fermi velocity along the stacks and $b$ and $t_{\perp}$ are the distance and transfer integral of adjacent stacks.
This approach is applicable only when transport is coherent along the chains ($\sigma_{\parallel}$) and incoherent in the perpendicular direction ($\sigma_{\perp}$).
Using tight binding approximation for a quarter-filled band, resulting in $\hbar v_{F}=\sqrt{2}t_{\parallel}a_s$, with $a_s$ equal to the distance of the molecules (unit-cell length $a=2a_s$), one ends up with:\cite{Soda77,Jerome82}
\begin{equation}
\label{eq:8}
\frac{\sigma_{\parallel}}{\sigma_{\perp}}=\left(\frac{a_s}{b}\right)^2 \left(\frac{t_{\parallel}}{t_{\perp}}\right)^2 \quad .
\end{equation}
In regard to (TMTTF)$_2$X, this relation is only valid within the temperature range $T>T_\rho$, and calculations for (TMTTF)$_2$PF$_6$ yield $1:16:860$ for the anisotropy, reproducing our experimental data (cf.\ Tab.~\ref{tab:1}) within a factor of 3.

Eq.~(\ref{eq:7}) can be extended to gapped system when replacing $v_F$ by the thermal velocity $v_T=\sqrt{k_BT/m^*}$:\cite{Ishiguro80}
\begin{equation}
\label{eq:9}
\frac{\sigma_{\parallel}}{\sigma_{\perp}}=\frac{\hbar^2 k_B T}{2 m^{\ast} b^2 t_{\perp}^2} \quad .
\end{equation}
This equation is only valid for very low carrier concentration, when phase space restrictions are not important any more and the Fermi distribution can be replaced by the Boltzmann distribution.
For (TMTTF)$_2$X and $T<T_\rho$ the overall temperature dependence of the measured anisotropy is in agreement with the Eq.~(\ref{eq:9}) for the gapped system.

These predictions assume perfect crystals and do not take defects and imperfections of real samples into account. Under extreme conditions, e.g.\ when the gap in the conduction band is very large, contributions from variable-range hopping between localized states may exceed band transport. This is reflected in our measurements, when for very low temperatures $\rho(T)$ deviates from Arrhenius law, changing to the temperature dependance of hopping transport.

\subsection{Luttinger liquid}

\begin{figure}
\centering\resizebox{\columnwidth}{!}{\includegraphics*{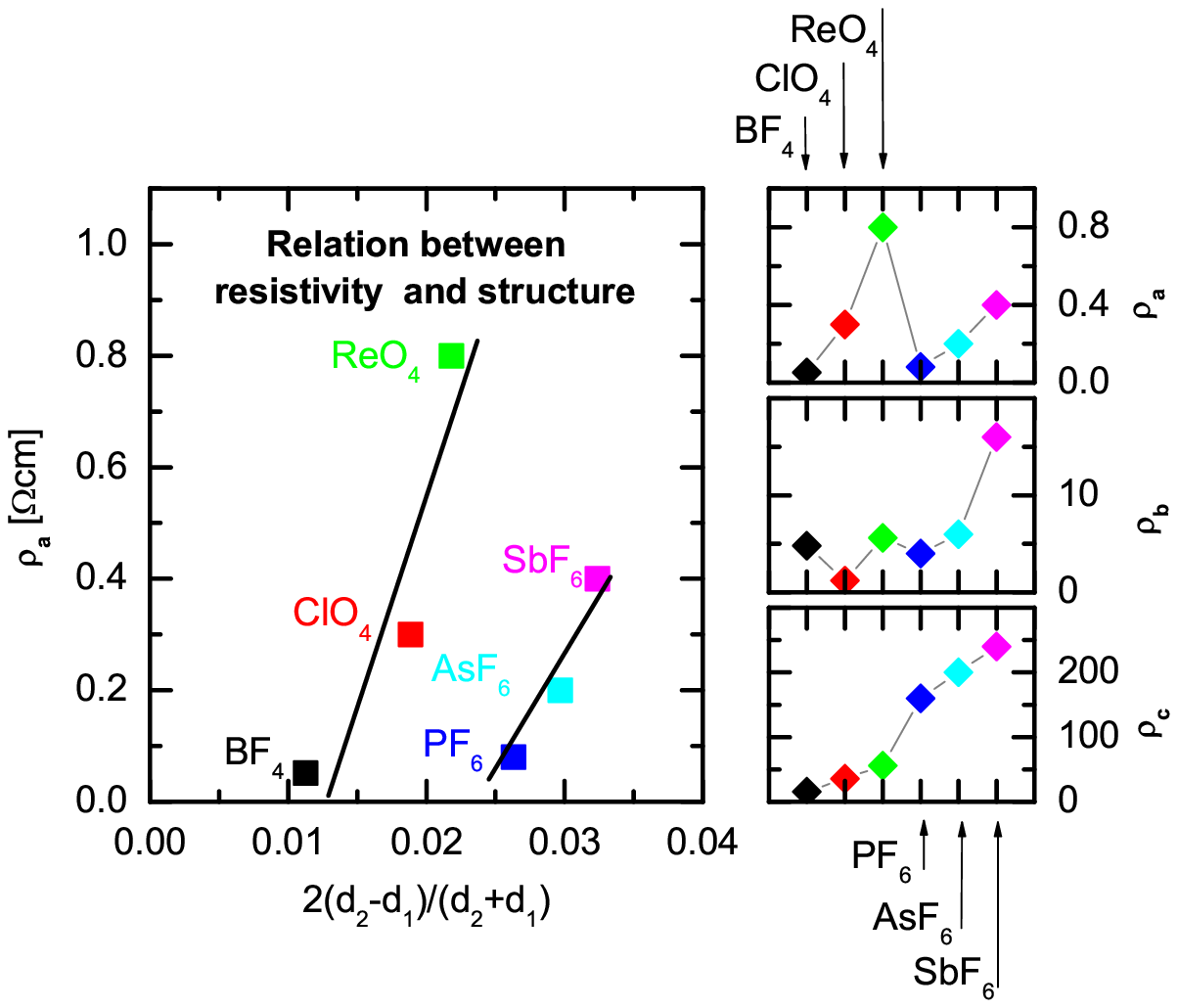}}
\caption{\label{fig:9} (Color online) In the left panel the dependence between the resistivity $\rho_a$ and the structural dimerization along $a$-axis, both at room temperature is plotted. The structural data are taken from Refs.~\onlinecite{Laversanne84,Liautard82,Galigne79,Iwase09,Liautard84,Kobayashi84}. The right panels show the room temperature resistivities $\rho_a$, $\rho_{b^{\prime}}$ and $\rho_{c^*}$ for all investigates compounds (TMTTF)$_2X$; from left to right the volume of the unit cell increases, {\it i.e.}\ $X$ = BF$_4$, ClO$_4$, ReO$_4$, PF$_6$, AsF$_6$, and SbF$_6$.}
\end{figure}

All investigated (TMTTF)$_2X$ salts are one-dimen\-sional conductors and located in the Luttinger-liquid re\-gime for high temperatures. When $T$ is lowered, electronic correlations become relevant and open up a gap $\Delta_0$ in the electronic density of states, {\it i.e.}\ drive the system into a Mott insulating state (cf.\ phase diagram Fig.~\ref{fig:8}). Due to the high localization temperature $T_\rho$ of about 250~K, it difficult to extract Luttinger exponents from our measurements. Transport studies at elevated  temperatures are highly desirable, however, the crystals decompose at approximately 400~K.

In the Bechgaard salts (TM\-TSF)$_2$\-$X$ a dimensional crossover is reported upon cooling or application of pressure, \cite{Jacobsen81,Vescoli98,Pashkin06,Pashkin10} for example seen in the development of a Drude-type conductivity and plasma edge in the optical properties of the $b$-direction. In the case of the sulfur analogues, such as (TMTTF)$_2$PF$_6$, pressure has to well exceed 20 kbar.\cite{Pashkin10} This is in full agreement with the (about a factor of 4) different  values for the transfer integrals $t_b$ of the  selenium and sulfur analogues, for instance (TMTSF)$_2$PF$_6$: $t_b\approx56$~meV and  (TMTTF)$_2$PF$_6$: $t_b\approx13$~meV.\cite{Bourbonnais09,Granier88}

\subsection{Charge localization}

The metal-like conduction at elevated temperatures vanishes around
250~K basically for all (TMTTF)$_2$$X$ compounds, and charge
localization sets in  due to the opening of a Mott gap. In the
range above the ordering temperature, transport is characterized
by an average activation energy $\Delta_0$ in the order of 500
$\pm$ 100 K for most compounds. With increasing anion size, the
energy $\Delta_0$ decreases; however, there seems to be no simple
relation to the dimerization observed in these compounds
(Tables~\ref{tab:3} and ~\ref{tab:4}). Williams {\it et al.}\cite{Williams85} pointed out a relation to the unit-cell volume and the S-S contacts which both increase when going from $X$ = BF$_4$ to ReO$_4$ and further to SbF$_6$ (Fig.~\ref{fig:11}).
\
Regarding the theoretical description of $\Delta_0$, different
factors have to be taken into account. The nominal band filling of
the (TMTTF)$_2X$ compounds is 1/4, but due to the structural
dimerization of periodicity $4k_F$ a band gap opens up, moving the system towards half filling.\cite{Emery83} Due to umklapp scattering, the system
develops the Mott gap $\Delta_0$. Following
Giamarchi\cite{Giamarchi97} the strength of the umklapp scattering
is described for half and quarter filled systems by
$g_\frac{1}{2}\propto{U}(D/E_{F})$ and
$g_\frac{1}{4}\propto{U}(U/E_{F})^2$ respectively, where $U$
stands for on-side Coulomb repulsion, $E_F$ is the Fermi energy
and $D$ denotes the dimerization gap. The strength of the
dimerization $D$ is difficult to estimate, since it is caused by
bond or on-site dimerization or by the anion potential.
Considering only bond dimerization along the $a$-direction, the
data for $\rho_{a}(T)$ show a linear relation within symmetric or
non-symmetric anions, plotted in Fig.~\ref{fig:9} as a function of
to the relative dimerization $2(d_2-d_1)/(d_2+d_1)$.

It is interesting to note, that for the perpendicular transport $\rho_{c^*}$, the ambient-temperature values increase
monotonously when going from $X$ = BF$_4$ to SbF$_6$ despite the
different symmetry (Fig.~\ref{fig:9}).

\subsection{Charge order}

\begin{figure}
\centering\resizebox{\columnwidth}{!}{\includegraphics*{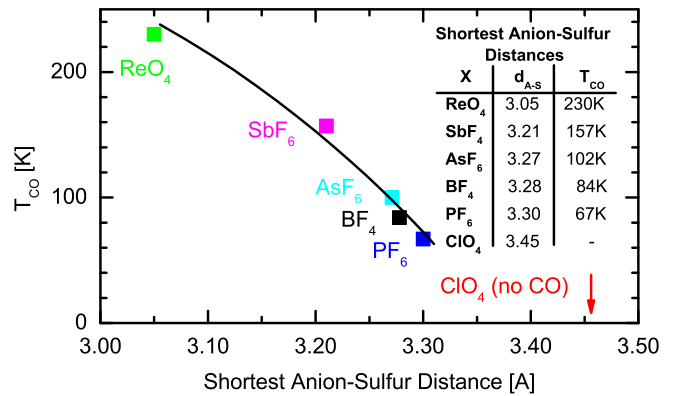}}
\caption{\label{fig:10} (Color online) The charge order transition temperature $T_{\rm CO}$ as a function of the shortest distance between the ligands of the anions (F or O) and sulfur atoms in TMTTF. For (TMTTF)$_2$ClO$_4$ the distance is much larger compared to the other compounds: it develops no charge order.
The structural data are taken from Refs.~\onlinecite{Laversanne84,Liautard82,Galigne79,Iwase09,Liautard84,Kobayashi84}.}
\end{figure}

Except (TMTTF)$_2$ClO$_4$, all Fabre salts develop a
charge-ordered phase below $T<T_{CO}$, as first observed in
thermopower and resistivity measurements. Interestingly, 
there is no decrease or distinct anomaly in the magnetic susceptibility:\cite{Salameh10} unlike common semiconductors, the spin is not affected by charge activation. The charge-ordered phase
was finally proven by NMR,\cite{Chow00,Zamborszky02}
dielectric\cite{Nad00,Monceau01,Nad06} and optical
measurements\cite{Dumm05,Hirose10}  and amounts to a few percent as listed
in the right columns of Tab.~\ref{tab:4}. 

The charge disproportionation results from an interplay of on-side and next-neighbor repulsion $U$ and $V$, respectively. While this phenomena is clear for quarter-filled compounds, it is still under discussion for systems close to half filling, which might be the case in (TMTTF)$_2X$ due to the structural dimerization of periodicity $4k_F$. It has been argued\cite{Takahashi06} that there is a strong competition between the dimeric Mott state and the charge-ordered state. According to Nogami \textit{et al.}\cite{Nogami05} the relative dimerization $\Delta t_s/\langle t_s \rangle
= 2(t_{s1} - t_{s2})/(t_{s1} + t_{s2})$
decreases strongly from its room-temperature value of $0.4 - 0.45$ when the temperature is reduced and reaches approximately 0.25 at $T_{\rm CO}$.  Values of $\Delta t_s/\langle t_s\rangle$ larger than 0.25 prevents $4k_F$ site charge order in (TMTTF)$_2X$. This can be interpreted, that compared to half-filled  properties quarter-filling has to dominate in these system in order to exhibit charge order. In Tab.~\ref{tab:4} the available values for dimerization and charge disproportionation are listed.
\begin{table}[b]
\caption{The relative dimerization $\Delta t_s/\langle t_s \rangle$ is determined by structural studies of different Fabre salts. The room temperature values are taken from Nogami {\it et al.},\cite{Nogami05} those in brackets are from Refs.~\onlinecite{Pouget96,Takahashi06,Ducasse86}. The charge order $2\delta$ characterizes the charge disproportionation between charge rich site $\rho_0 + \delta$ and charge poor sites $\rho_0 - \delta$ as derived from optical and magnetic measurements (Refs.~\onlinecite{Dumm05,Hirose10,Nakamura07,Nogami02,Takahashi06,Chow00,Zamborszky02}) conducted at different temperatures as indicated in brackets.
\label{tab:4}}
\begin{center}
\begin{tabular}{c|c c|c c}
\hline\noalign{\smallskip}
 \multicolumn{4}{c}{~~~~~~~~(TMTTF)$_2X$}\\
\noalign{\smallskip}\hline\noalign{\smallskip}
$X$ & \multicolumn{2}{c|}{$\Delta t_s/\langle t_s \rangle$} & \multicolumn{2}{c}{$\delta$} \\
&$T\approx 300$~K&low $T$& NMR& IR\\
\noalign{\smallskip}\hline\noalign{\smallskip}
PF$_6$  &~0.41(0.38)&~0.20~(30K)&~~~0.12(30K)&~0.06(30K)\\
AsF$_6$ &~0.46(0.34)&~0.19~(40K)&~~~0.13(30K)&~0.13(20K)\\
SbF$_6$ &~0.39(0.30)&~0.17~(100K)&~~~0.25(low)&~\\
ClO$_4$ &~0.34(0.33)& &~-~~&~-\\
ReO$_4$ &~0.45~~&~0.30~(150K) &~0.17(100K)&~
\\\noalign{\smallskip}\hline
\end{tabular}
\end{center}
\end{table}

For a along time, the charge-order transition in (TMTTF)$_2X$ was considered of purely electronic origin and called structureless transition.\cite{Coulon85} Since no additional spots could be found in x-ray
scattering,\cite{Nogami02,Nogami05} the possible structural changes had to be very small.
Only recently, high resolution thermal expansion measurements\cite{DeSouza08} revealed a pronounced anomaly in (TMTTF)$_2$PF$_6$ and (TMTTF)$_2$AsF$_6$ at the charge-order transition, that is much stronger pronounced along $c^{\ast}$ and $b^{\prime}$-axes than along the chains. These results support early suggestions of Pouget \textit{et al.},\cite{PougetBookJerome} and others\cite{Brazovskii08} that charge order is stabilized by structural modifications, in particular by the anions.\cite{Foury10}

The fact that charge order is linked to the anion arrangement becomes obvious from Fig.~\ref{fig:10}: a linear correlation is discovered when the charge-order transition temperature $T_{\rm CO}$ is plotted versus the shortest distance between the sulfur atoms in the TMTTF molecule and the ligands of the anions, {\it i.e.}\ fluorine or oxygen.
Note, the only compound that does not develop charge disproportionation by far holds the longest distance.\cite{remark3} The important interaction between the organic cations and the counterions was recently detected by ESR spectroscopy which revealed an anomalous temperature behavior of the $g$-factor, indicating the deformation of the molecular orbitals by the anion potential.\cite{Furukawa09}

By comparing the shape of the anions of the different compounds, it is obvious, that especially the non-symmetric anions are strongly distorted; this calls for a closer inspection of the anions themselves.
A short contact between ligand and organic molecule is naturally correlated with a long bond between the ligand and the central atom of the anion. The bond length in the anions increases from ClO$_4$ over BF$_4$ to ReO$_4$. This is determined by the size of the central atom and the difference in electronegativity of the atoms constituting the anion.

Despite of some ambiguity in extracting the slope in the Arrhenius
plot,\cite{remark2} the energy gaps deduce from our transport
measurements (Tab.~\ref{tab:3}) are in fair agreement with
literature data. In general the activation energy increases upon
lowering the temperature from one regime to another. Below $T_{\rm CO}$ the charge order gap $\Delta_{\rm CO}$(T) following a BCS-like
behavior, indicating a second-order phase transition. Since we
start out from a charge-localized state, it adds to the present
gap $\Delta_0$ according to Eq.~(\ref{eq:gapquadrat}). The total
activation energy is a consequence of bond and site contributions to the
umklapp scattering, {\it i.e.}\ $\Delta=\Delta(U)$ and
$U=\sqrt{{U_b}^2+{U_s}^2}$.\cite{Brazovskii81,Brazovskii08} Our assumption to add up bond gap and the side gap according to Eq.~(\ref{eq:gapquadrat}) is not mandatory for pure electronic origin, but the good fit to our data justifies this assumption. It can be seen as an indication that the charge-order gap does not only have electronic origin,
but is stabilized by structural properties. It should be noted
that significant differences exist for the various compounds, and
also the increase of the activation energy upon passing through
the CO transition varies significantly. Thus it is not possible to
related $\Delta_{\rm CO}(T)$ to the value of the charge order
parameter $\delta$ used to characterize the charge
disproportionation.

\begin{figure}
\centering\resizebox{0.8\columnwidth}{!}{\includegraphics*{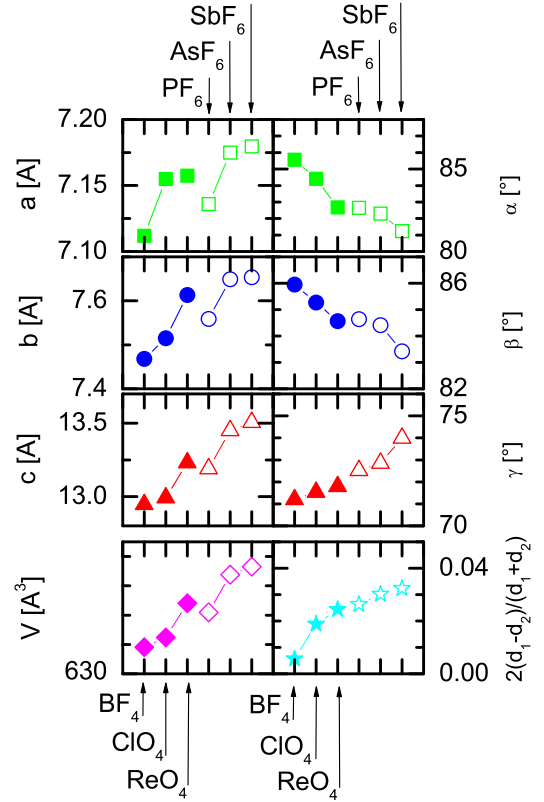}}
\caption{\label{fig:11}(Color online)  Unit cell and dimerization parameters for the six investigated compounds (TMTTF)$_2X$ . The compounds are ordered form left to right by increasing unit cell volume ($X$ = BF$_4$, ClO$_4$, ReO$_4$,  PF$_6$, AsF$_6$, and SbF$_6$). The structural data are taken from Refs.~\onlinecite{Laversanne84,Liautard82,Galigne79,Iwase09,Liautard84,Kobayashi84}
and tabulated in footnote~\onlinecite{remark5}.}
\end{figure}

As depicted in Fig.~\ref{fig:11},  there seem to be some
correlations between the various anions and structural parameters.
However, due to the different symmetry and distortion of the
anions, a rigorous relation of unit-cell parameters with physical
properties seems difficult. Further issues of relevance are the
polarizibility and how strong charge is located on the anion.

One important finding of our study is the surprising  similarity
of the overall behavior in all three directions. This implies that
the charge ordering is coupled between the chains and influences
the transport in a similar way. Brazovskii pointed out that in the
present case of a ferroelectric Mott-Hubbard insulator ordered
domains develop, separated by ferroelectric domain
walls.\cite{Brazovskii08} Further experimental and theoretical
studies of the behavior in the perpendicular directions are highly desirable.

\subsection{Anion order}

As demonstrated from Fig.~\ref{fig:3}, the anion-order anomaly in (TM\-TTF)$_2$\-ClO$_4$ is not associated with a noticeable change
of the slope in resistivity $\rho(T)$, as already pointed out by
Coulon {\it et al.} \cite{Coulon82} Finding the identical
activation energy on both sides of the phase transition implies
that the dominant transport mechanism remains unchanged. In the case of (TMTTF)$_2$\-BF$_4$, a slight
increase of the activation energy can be identified when passing
$T_{\rm AO}$. The maximum anion order gap $\Delta_{\rm AO}$ is found for (TMTTF)$_2$\-ReO$_4$, following a BCS-like temperature dependence. 

For $X$ = BF$_4$ and ClO$_4$ a step-like reduction in resistivity at $T_{\rm AO}$ is observed. It can be interpreted as a freezing out of scattering channels 
when anions order. The fact that the overall slope $\rho(T)$ does not change significantly above and below $T_{\rm AO}$ implies that  the more or less temperature independent scattering due to anion disorder --~which adds to temperature-dependent factors according to Matthiessen's rule~-- just freezes out.

Taking into account these different characteristics of the anion order in the investigated (TMTTF)X$_2$ compounds, the conclusion is obvious, that $\Delta_{\rm AO}$ is not only generated by backscattering on the anion potential but also by changes which the anion order induces on the TMTTF-molecules. The value of $\Delta_{\rm AO}$ depends, however, on the coupling between anion and TMTTF-stack. The fact that in (TMTTF)$_2$SCN the metal-insulator transition occurs at $T_{\rm AO}$ and the anion superstructure does not lead to a doubling of the unit cell along the $a$-direction [\,$q$=(0\,,\,1/2\,,\,1/2), Ref.~\onlinecite{coulon82ab}] supports this idea. Recent ESR experiments corroborate this picture as well:\cite{Salameh10} The anion ordering in
(TMTTF)$_2$ReO$_4$ leads to a very large singlet-triplet gap
$\Delta_{\sigma}=1100$~K because a \,o\,-\,O\,-\,O\,-\,o\,- charge
pattern develops along the stacks below $T_{\rm AO}$.

Coupling strength between anions and TMTTF-stack seems not only to depend on the size of the anion, increases from BF$_4$,ClO$_4$ to ReO$_4$, even more pronounced is the difference in the shortest distance from anions-ligand to sulfur (tabulated in the inset of Fig.~\ref{fig:10}) and the electronegativity of the atoms constituting the anions. Following this idea, coupling increases from ClO$_4$, BF$_4$ to ReO$_4$, whereas the transition temperature depends on the anion size.

The step in $\rho(T)$ is the same in all three directions, implying that the ordering affects the transport in a similar way. We conclude  that a three-dimensional order of the anions takes place at $T_{\rm AO}$; {\it i.e.}\ there is strong coupling between the stacks and the periodicity changes in all directions. These findings are in accord with previous and recent ESR experiments.\cite{Dumm00,Salameh10}

\section{Summary}
The temperature dependent dc resistivity of the
quasi-one-dimensional organic salts (TMTTF)$_2X$ ($X$ = PF$_6$,
AsF$_6$, SbF$_6$; BF$_4$, ClO$_4$, ReO$_4$) was measured and
analyzed in all three directions. Most of the compounds exhibit a weak metallic behavior at elevated temperatures. Increasing charge localization leads to a resistivity minimum around $T_{\rho}=250$~K and the opening of an energy gap which upon cooling increases up to $\Delta_0\approx 400$~K before charge order sets in.
Below $T_{\rm CO}$ we can derive a gap $\Delta_{\rm CO}(T)$ in the density of states that opens in a mean-field-like behavior. The effect is seen in all three crystal directions, but best resolved along $c^{*}$ axis.
The resistivity increases in all compounds, but to a different extend: in (TMTTF)$_2$PF$_6$ it is barely visible, the effect of charge order is moderate in (TMTTF)$_2$BF$_4$ and
AsF$_6$, and very strong in (TMTTF)$_2$SbF$_6$ and
(TM\-TTF)$_2$\-ReO$_4$. From our comparison of structural and transport data, we find relations of resistivity values to the anion size and amount of dimerization; also we can clearly identify the influence of the anions on the charge-order transition.
Anion ordering  in
(TMTTF)$_2$BF$_4$ and (TM\-TTF)$_2$\-ClO$_4$ causes a step-like
decrease of the resistivity below $T_{\rm AO}$ since disorder scattering on reduced.  Contrary, the energy gap of (TMTTF)$_2$\-ReO$_4$ increases strongly because of the much stronger influence of the anion on the organic stacks, leading to a variation of the charge pattern. The shape of the anion-order anomaly seems to be related to the difference in the coupling between anions and TMTTF-stacks.

\begin{acknowledgements}
We thank B.W. Fravel, L. Montgomery and D. Schweitzer for providing
the  TMTTF molecules for our crystal growth. Valuable discussions with and numerous comments of S.A. Brazovskii, S. Brown, A. Greco, T. Ishiguro, D. J\'{e}rome, J. Merino, and J.-P. Pouget are appreciated. The work was supported by the Deutsche Forschungsgemeinschaft.
\end{acknowledgements}


\begin{thebibliography}{99}
\bibitem{Kagoshima88}
S. Kagoshima, H. Nagasawa, and T. Sambongi, {\em One-dimensional conductors} (Springer, Berlin, 1988).
\bibitem{Dressel03}
M. Dressel, Naturwissenschaften {\bf 90}, 337 (2003).
\bibitem{Dressel07}
M. Dressel, Naturwissenschaften {\bf 94}, 527 (2007).
\bibitem{Jerome82}
D. J\'{e}rome and H. J. Schulz, Adv. Phys. {\bf 31},  299  (1982).
\bibitem{Jerome94}D. J\'erome,  in {\em Organic Conductors}, edited by J.-P. Farges (Marcel
Dekker, New York, 1994), p.\ 405.
\bibitem{Ishiguro98}T. Ishiguro, K. Yamaji, and G. Saito, {\em Organic
Superconductors}, 2nd edition (Springer-Verlag, Berlin, 1998).
\bibitem{Pouget96}
J. P. Pouget and S. Ravy, J. Phys. I (France) {\bf 6}, 1501 (1996).
\bibitem{Brazovskii08}
S.A. Brazovskii, {\em Ferroelectricity and Charge Ordering in
Quasi-1D Organic Conductors}, in: {\em The Physics of Organic
Superconductors and Conductors}, edited by A.G. Lebed
(Springer-Verlag, Berlin, 2008) , p. 313.
\bibitem{Dressel01}
M. Dressel, P. Hesse, S. Kirchner, G. Untereiner, M. Dumm, J. Hemberger,
A. Loidl, and L. Montgomery, Synth. Met. {\bf 120}, 719 (2001).
\bibitem{Korin06}
B. Korin-Hamzi\'c, E. Tafra, M. Basleti\'c, A. Hamzi\'c, and M. Dressel, Phys. Rev. B {\bf 73}, 115102 (2006).
\bibitem{Coulon82}
C. Coulon, P. Delhaes, S. Flandrois, R. Lagnier, E. Bonjour, and J.M. Fabre, J. Phys. (France) {\bf 43}, 1059 (1982).
\bibitem{Moret83}
R. Moret, J. P. Pouget, R. Comes, and K. Bechgaard, J. Phys. Colloq. (France) {\bf 44} C3-957 (1983).
\bibitem{Pouget82}
J.P. Pouget, R. Moret, R. Comes, K. Bechgaard, J. M. Fabre, and L. Giral, Mol. Cryst. Liq. Cryst. {\bf 79}, 129 (1982).
\bibitem{Ravy86}
S. Ravy, R. Moret, J. P. Pouget, and R. Comes, Synth. Met. {\bf 13}, 63 (1986).
\bibitem{Nad06}
F. Nad and P. Monceau, J. Phys. Soc. Jpn. {\bf 75}, 051005 (2006)
\bibitem{Mihaly00}
G. Mih\'{a}ly, I.  K\'{e}zsm\'{a}rki, F. Z\'{a}mborszky, and L. Forr\'{o}, Phys. Rev. Lett. {\bf 84}, 2670 (2000).
\bibitem{Rose10}
E. Rose, M. Dumm, and M. Dressel, to be published.
\bibitem{Mott79}
N. F. Mott and E. Davies, {\em Electronic Processes in Non-Crystalline Materials}, (Oxford University Press, Oxford, 1979).

\bibitem{remark1}
The values of $\rho_0(T)$ in ($\rm\Omega$cm) are as follows:\\
\begin{tabular}{l|ccc}
compound&$a$ axis & $b$ axis  & $c$ axis\\
\hline
(TMTTF)$_2$PF$_6$&0.068 & 0.33  & 15\\
(TMTTF)$_2$As$_6$&0.068 & 0.55 & 9.5\\
(TMTTF)$_2$Sb$_6$&0.229 & 0.9  & 58\\
(TMTTF)$_2$BF$_4$&0.047 & 1.37  & 5.6\\
(TMTTF)$_2$ClO$_4$&0.19 & 0.25  & 5\\
(TMTTF)$_2$ReO$_4$&0.21 & 0.95  & 7.3\\
\end{tabular}

\bibitem{Takahashi06}
T. Takahashi, Y. Nogami, and K. Yakushi, J. Phys. Soc. Jpn. {\bf 75}, 051008 (2006).

\bibitem{Georges00}
A. Georges, T. Giamarchi, and N. Sandler, Phys. Rev. B {\bf 61}, 16393 (2000).
\bibitem{remark4}
Note, that is also compatible with an expectation of Brazovskii \protect\cite{Brazovskii08} that the transport is carried out by (spinless) solitons, {\it i.e.}\ holons. As topological objects, the solitons need pair collisions or annihilations to  travel between chains.
\bibitem{Galigne79}
J. L. Galign\'{e}, B. Liautard, S. Peytavin, G. Brun, M. Maurin, J.M. Fabre, E. Torreilles and L. Giral, Acta Cryst. B {\bf35}, 1129 (1979).
\bibitem{Liautard84}
B. Liautard, S. Peytavin, G. Brun, D. Chasseau, J.M. Fabre and L. Giral, Acta Cryst. C {\bf40}, 1023 (1984).
\bibitem{Kobayashi84}
H. Kobayashi, A. Kobayashi, Y. Sasaki, G. Saito and H. Inokuchi, Bull. Chem. Soc. Jpn. {\bf57}, 2025 (1984); There seems to be a typing error in the unit cell data: 
$\alpha=82.66^\circ$ (instead of $86.62^\circ$) is in agreement with other measurements (e.g. Ref.~\onlinecite{Furukawa05}) and fits to the unit cell volume of $V=679.5$~\AA$^3$ published in the article. The stacking distances of the molecules along a-direction extracted from the corresponding \textit{cif}-file are: $d_1=3.615$~\AA\ and $d_2=3.537$~\AA\ and thus differ from those mentioned in the publication.
\bibitem{Iwase09}
F. Iwase, K. Sugiura, K. Furukawa and T. Nakamura, J. Phys. Soc. Jpn. {\bf 78}, 104717 (2009).
\bibitem{Liautard82}
B. Liautard, S. Peytavin, G. Brun and M. Maurin, Cryst. Struc. Comm. C {\bf11}, 1841 (1982).
\bibitem{Furukawa05}
K. Furukawa, T. Hara and T. Nakamura, J. Phys. Soc. Jpn. {\bf 74}, 3288 (2005).
\bibitem{Laversanne84}
R. Laversanne, C. Coulon, B. Gallois, J.P. Pouget and R. Moret, J. Physique Lett. {\bf45}, L-393 (1984).
\bibitem{Ishiguro80}
T. Ishiguro, H. Sumi, S. Kagoshima, K. Kajimura and H. Anzai, JPSJ {\bf 48}, 456 (1980).
\bibitem{Soda77}
G. Soda, D. J\'{e}rome, M. Weger, J. Alizon, J. Gallice, H. Robert, J. M. Fabre, and L. Giral, J. Phys. (France) {\bf 38}, 931 (1977).
\bibitem{Jacobsen81}
C. S. Jacobsen, D. B. Tanner, and K. Bechgaard, Phys. Rev. Lett.
{\bf 46}, 1142 (1981).
\bibitem{Vescoli98}
V. Vescoli, L. Degiorgi, W. Henderson, G. Gr\"uner, K. P. Starkey,
and L. K. Montgomery, Science {\bf 281}, 1181 (1998).
\bibitem{Pashkin06}
A. Pashkin, M. Dressel, and C. A. Kuntscher, Phys. Rev. B {\bf 74},
165118 (2006).
\bibitem{Pashkin10} A. Pashkin, M. Dressel, M. Hanfland, and C. A. Kuntscher, Phys. Rev. B {\bf 81}, 125109 (2010).

\bibitem{remark5}
Structural data of the investigated (TMTTF)$_2X$ compounds. The references of the different compounds are in sequence \onlinecite{Galigne79,Liautard84,Kobayashi84,Iwase09,Liautard82} and \onlinecite{Iwase09}, data indicated by (*) are extracted from the associated cif-file. The values for the dimerization in (TMTTF)$_2$PF$_6$ and (TMTTF)$_2$SbF$_6$ are out of Ref.~\onlinecite{Laversanne84}.

\begin{tabular}{l|l l l l}
\noalign{\smallskip}
$~~X$ & $~~a$ (\AA) &  $~~b$ (\AA)& $~~c$ (\AA)& $~\Delta d_a/\langle d_a \rangle$ \\
\noalign{\smallskip}\hline\noalign{\smallskip}
BF$_4$ &~7.112&~7.468&~12.946&~~$1.13\times 10^{-2 \ (*)}$\\
ClO$_4$  &~7.155 &~7.515&~12.992&~~$1.89\times 10^{-2 \ (*)}$\\
ReO$_4$ &~7.158~~&~7.613&~13.231&~~$2.10\times 10^{-2 \ (*)}$\\
PF$_6$ &~7.157&~7.580&~13.213&~~$2.63\times 10^{-2}$\\
AsF$_6$ &~7.178&~7.610&~13.317&~~$2.97\times 10^{-2 \ (*)}$\\
SbF$_6$ &~7.180&~7.654 &~13.507&~~$3.24\times 10^{-2}$\\
\noalign{\smallskip}\hline\noalign{\smallskip}
 &~~$\alpha$ ($^\circ$) &~~$\beta$ ($^\circ$)&~~$\gamma$ ($^\circ$)&~~$V$ (\AA$^3$) \\
 \noalign{\smallskip}\hline\noalign{\smallskip}
 BF$_4$ &~85.56&~85.95&~71.20&~~648.2\\
ClO$_4$  &~84.41 &~85.27&~71.53&~~648.8\\
ReO$_4$ &~82.68~~&~84.57&~71.79&~~678.2\\
PF$_6$ &~82.64&~84.72&~72.41&~~676.6\\
AsF$_6$ &~82.03&~84.25&~72.89&~~697.7\\
SbF$_6$ &~81.24&~83.42 &~74.00&~~702.9\\
\noalign{\smallskip}\hline
\end{tabular}

\bibitem{Granier88}
T. Granier and B. Gallois, L. Ducasse, A. Fritsch, and A. Filhol, Synthetic Metals {\bf 24}, 343 (1988)
\bibitem{Bourbonnais09}
C. Bourbonnais and D. Jerome, Cond. Mat., 0904.0617 (2009)
\bibitem{Williams85}
J. M. Williams, M. A. Beno, H.-H. Wang, T. J. Emge, P. T. Copps,
L. N. Hall, K. D. Carlson, and G. W. Crabtree, Phil. Trans. Royal Soc. (London) A {\bf 314}, 83 (1985).
\bibitem{Emery83}
V. J. Emery, J. Phys. Colloq. (France) {\bf 44} C3-977 (1983).
\bibitem{Giamarchi97}
T. Giamarchi, Physica B {\bf 230-232}, 975 (1997); T. Giamarchi, {\em From Luttinger to Fermi Liquids in Organic Conductors}, in: {\em The Physics of Organic Superconductors and Conductors}, edited by A. Lebed, (Springer-Verlag, Berlin 2007), p.~719.
\bibitem{Salameh10}
B. Salameh, S. Yasin, M. Dumm, M. Untereiner, L. Montgomery, and M. Dressel, submitted to Phys. Rev. B.
\bibitem{coulon82ab}
C. Coulon, A. Maaroufi, J. Amiell, E. Dupart, S. Flandrois, P. Delhaes, R. Moret, J.P. Pouget, and J.P. Morand, Phys. Rev. B {\bf 26}, 6322 (1982); C. Coulon, P. Delhaes, S. Flandrois, R. Lagnier, E. Bonjour, and J.M. Fabre, J. Phys. (France) {\bf 43}, 1059 (1982). 
\bibitem{Chow00}
D.S. Chow, F. Zamborszky, B. Alavi, D.J. Tantillo, A. Baur, C.A. Merlic, and S.E. Brown, Phys. Rev. Lett. {\bf 85}, 1698 (2000).
\bibitem{Zamborszky02}
F. Zamborszky, W. Yu, W. Raas, S.E. Brown, B. Alavi, C.A. Merlic, and A. Baur, Phys. Rev. B {\bf 66}, 081103 (2002)
\bibitem{Nad00}F.Y. Nad, P. Monceau, C. Carcel, and J.M. Fabre, Phys. Rev. B {\bf 62}, 1753 (2000).
\bibitem{Monceau01}P. Monceau, F.Y. Nad, and S. Brazovskii, Phys. Rev. Lett. {\bf 86}, 4080 (2001).
\bibitem{Dumm05}
M. Dumm, M. Abaker, and M. Dressel, J. Phys. IV (France) {\bf 131}, 55 (2005); M. Dumm, M. Abaker, M. Dressel, and L. K. Montgomery, J. Low Temp. Phys. {\bf 142}, 609 (2006).
\bibitem{Hirose10}
S. Hirose, A. Kawamoto, N. Matsunaga, K. Nomura,
K. Yamamoto and K. Yakushi, Phys. Rev. B {\bf 81}, 205107  (2010).
\bibitem{Nakamura07}
T. Nakamura, K. Furukawa, and T. Hara, J. Phys. Soc. Jpn. {\bf 76}, 064715 (2007).
\bibitem{Nogami02}
Y. Nogami and T. Nakamura, J. Phys. IV (France) {\bf 12}, Pr-9-145 (2002).
\bibitem{Nogami05}
Y. Nogami, T. Ito, K. Yamamoto, N. Irie, S. Horita, T. Kambe, N. Nagao, K. Oshima, N. Ikeda, and T. Nakamura, J. Phys. IV (France) {\bf 131}, 39 (2005).
\bibitem{Ducasse86}
L. Ducasse, M. Abderrabba, J. Hoarau, M. Pesquer, B. Gallois, and J. Gaultier, J. Phys. C {\bf 19}, 3805 (1986); L. Ducasse, M. Abderrabba, B. Gallois, and D. Chasseau, Synth. Metals {\bf 19}, 327 (1987).
\bibitem{Coulon85}
C. Coulon, S. S. P. Parkin, and R. Laversanne, Phys. Rev. B {\bf 31}, 3583 (1985).
\bibitem{DeSouza08}
M. de Souza, P. Foury-Leylekian, A. Moradpour, J. P. Pouget, and M. Lang, Phys. Rev. Lett. {\bf 101}, 216403 (2008);
M. de Souza, A. Br\"uhl, J. M\"uller, P. Foury-Leylekian, A. Moradpour, J.-P. Pouget, and M. Lang, Physica B {\bf 404}, 494 (2009);
M. de Souza, D. Hofmann, P. Foury-Leylekian, A. Moradpour, J. P. Pouget, and M. Lang, Physica B {\bf 405}, Suppl. 1, S92 (2010).
\bibitem{PougetBookJerome}
D. P. Pouget in {\em Low-Dimensional Conductors and Superconductors}, edited by D. Jerome and L. G. Caron (NATO ASI Series,Series B: Physics Vol. 155, New York, 1986), p.\ 38.
\bibitem{Foury10}
P. Foury-Leylekian, S. Petit, G. Andre, A. Moradpour, and J.-P. Pouget,
Physica B {\bf 405}, Suppl. 1, S95 (2010).
\bibitem{Furukawa09}
K. Furukawa, T. Hara, and T. Nakamura, J. Phys. Soc. Jpn. {\bf 78}, 104713 (2009).
\bibitem{remark3}
It is a particular fortune that for  (TMTTF)$_2$$X$ 
the more common $4k_F$ instability (charge order) coincides 
with the cage-like setting for the anions. The latter is prone to Earnshaw's instability of all classical charges; for small anions it may not be stabilized by rigidity of the methyl groups constituting the cage.
\bibitem{remark2}
Sometimes the extracted energy gaps deviate beyond the error bars; this  can only partially be attributed to different sample and contact quality.
\bibitem{Brazovskii81}
S.A. Brazovskii and N.N. Kirova, JETP Lett. {\bf 33}, 4 (1981).
\bibitem{Dumm00}
M. Dumm,  A. Loidl,  B. W. Fravel, K. P. Starkey,  L. K. Montgomery, and M. Dressel, Phys.\ Rev.\ B {\bf 61},  511 (2000); M. Dumm, B. Salameh, M. Abaker, L. K. Montgomery, and M. Dressel, J. Phys. IV (France) {\bf 114}, 57 (2004).


\end{thebibliography}
\end{document}